\documentclass[11pt,twoside]{article}
\usepackage{asp2010}

\resetcounters

\begin{document}

\title{Variability of Mini-BAL and BAL Outflows in Quasars}
\author{Paola Rodr{\'i}guez Hidalgo$^{1}$, Fred Hamann$^2$, Michael Eracleous$^1$, Daniel Capellupo$^2$, Jane Charlton$^1$, and Joseph Shields$^3$}
\affil{$^1$Department of Astronomy \& Astrophysics, The Pennsylvania State University, University Park, PA 16802}
\affil{$^2$Department of Astronomy, University of Florida, Gainesville, FL 32611}
\affil{$^3$Department of Physics \& Astronomy, Ohio University, Athens, OH 45701 }

\begin{abstract}

We report the results of several programs to study the variability of high-velocity (up to 0.2c) mini-"broad absorption lines" (mini-BALs) and BALs in quasar spectra, and thus to better characterize the structural and physical properties of these outflows. After the report of a highly variable mini-BAL outflow at a speed of $\sim$~0.17c in the quasar PG0935+417, we created the first systematic accounting of outflows in Sloan Digital Sky Survey (SDSS) quasar spectra that includes mini-BALs and extremely high velocity outflows (up to 0.2c) to measure their frequency. Following this study, we began a monitoring campaign to study the location, and dynamical and evolutionary effects of these outflows. This program covers a range of 0.9--3.3 years in the quasars' rest-frame by comparing new spectra (using facilities at the Kitt Peak National Observatory and MDM Observatory) with archival SDSS spectra. We find that ~57\% of quasars with mini-BALs and BALs varied between just two observations. This variability tends to occur in complex ways; however, all the variable lines vary in intensity and not in velocity, not finding evidence for acceleration/deceleration in these outßows. Due to the variations in strength, mini-BALs can become BALs and viceversa, suggesting they share a similar nature. We include as an example the discovery of the transition of a mini-BAL into a BAL in the spectra of the SDSS quasar J115122+020426.
\end{abstract}

\section{Introduction}

Outflows are fundamental constituents of Active Galactic Nuclei (AGN). They are commonly detected and might be ubiquitous if the absorbing gas subtends only part of the sky as seen from the central continuum source. Previously the most studied classes of outflows have been the``broad absorption lines", or BALs, and the "narrow absorption lines", or NALs (i.e., \citealt{Weymann81}; \citealt{Turnshek84};  \citealt{Foltz86}; \citealt{Aldcroft94}; \citealt{Reichard03}; \citealt{Vestergaard03}; \citealt{Trump06}). More recently, an intermediate class of mini-BALs and extremely high velocity outflows (up to 0.2c speeds) are increasing the percentage of observed outflows in quasars. Studies of the variability of these absorption systems help with our understanding of whether different subclasses represent the same or similar types of absorbers. Moreover, characterizing their variability can help locate these systems with respect to the inner quasar region and describe their physical structure (see Proga et al.~in these proceedings). 

\section{Variability of Mini-BALs and BALs}
\label{sec2}

One of the first found cases of a {{\rm C~}\kern 0.1em{\sc iv}~$\lambda\lambda 1548, 1550$} mini-BAL at extremely high velocities ($\sim$~) was reported by \citet{Hamann97b}. The formation of these absorbers, which appear as absorption lines in the quasar spectra of PG0935+417, in an extreme high-velocity quasar outflow (with $v \sim -50000$ km~s$^{-1}$) is confirmed by the line variability (\citealt{Narayanan04}; \citealt{RodriguezHidalgo10a}), broad smooth absorption profiles, and partial covering of the background light source. The line profiles are complex and asymmetric, with Full Widths at Half Minimum (FWHMs) of different components in the range $\sim$~660 to $\sim$~2510 km~s$^{-1}$. In \citet{RodriguezHidalgo10a} we report the detection of {{\rm O}\kern 0.1em{\sc vi}~$\lambda\lambda 1031, 1037$} and {{\rm N}\kern 0.1em{\sc v}~$\lambda\lambda 1238, 1242$} absorption, found in HST/FOS spectra, at the same redshift as the {\hbox{{\rm C~}\kern 0.1em{\sc iv}}} system. The resolved {\hbox{{\rm O}\kern 0.1em{\sc vi}}} doublet indicates that these lines are moderately saturated, with the absorber covering $\sim$~80\% of the quasar continuum source ($C_f \sim$~0.8). We derive ionic column densities of order 10$^{15}$ cm$^{-2}$ in {\hbox{{\rm C~}\kern 0.1em{\sc iv}}} and several times larger in {\hbox{{\rm O}\kern 0.1em{\sc vi}}}, indicating an ionization parameter of $\log U {\mathrel{\rlap{\lower4pt\hbox{\hskip1pt$\sim$}} \raise1pt\hbox{$>$}}} -0.5$. Assuming solar abundances, we estimate a total column density of $N_H \sim 5 \times 10^{19}$ cm$^{-2}$. 

\begin{figure}
\begin{center}
\includegraphics[width=10.5cm]{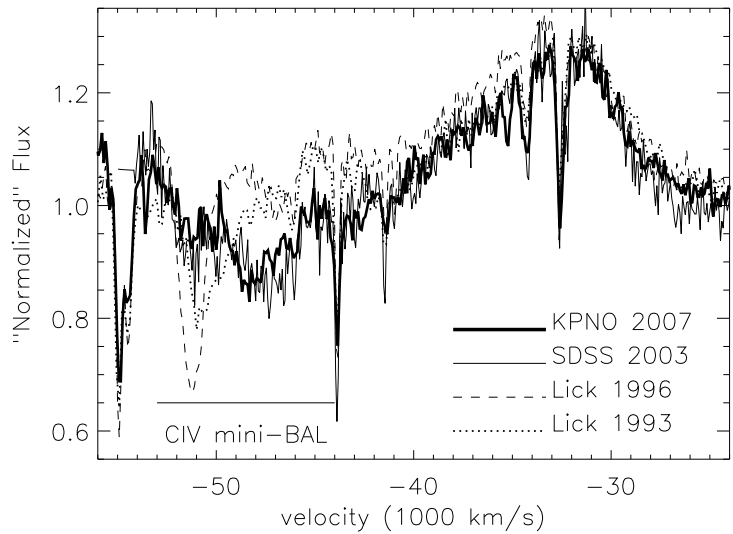}
\caption[Figure1]{Variability in the high velocity {{\rm C~}\kern 0.1em{\sc iv}~$\lambda\lambda 1548, 1550$} feature in the spectrum of the PG0935+417 quasar over a period of fourteen years (4.7 years in the quasar restframe). Spectra around the absorption troughs are shown after dividing by a linear fit to the continuum to highlight variations in absorption. The absorption at $\sim$~$-51000~$ {\hbox{km~s$^{-1}$}} in the SDSS and KPNO spectra is weaker than in previous measurements. However, the component outflowing at a slightly lower velocity ($\sim-47000$ km s$^{-1}$) seems stronger and wider in the two more recent spectra.}
\label{fig1}
\end{center}
\end{figure}

Comparisons to data in the literature show that this outflow emerged sometime between 1982 when it was clearly not present \citep{Bechtold84} and 1993 when it was first detected \citep{Hamann97b}. Our examination of the {\hbox{{\rm C~}\kern 0.1em{\sc iv}}} data from 1993 to 2007 shows that there is variable complex absorption across a range of velocities from $-45000$ to $-54000$ {\hbox{km~s$^{-1}$}} (see Figure \ref{fig1}). There is no clear evidence for acceleration or deceleration of the outflow gas. The observed line variations appear to occur in timescales of ${\mathrel{\rlap{\lower4pt\hbox{\hskip1pt$\sim$}} \raise1pt\hbox{$<$}}}$ ~1 year in the quasar rest-frame, which agrees with current theoretical results (see Proga et al. in these proceedings). These variations are consistent with either changes in the ionization state of the gas or clouds crossing our lines of sight to the continuum source. If the former case, the recombination times constrain the location of outflow to be at a radial distance of $r {\mathrel{\rlap{\lower4pt\hbox{\hskip1pt$\sim$}} \raise1pt\hbox{$<$}}}$~1.2 kpc with density of $n_H {\mathrel{\rlap{\lower4pt\hbox{\hskip1pt$\sim$}} \raise1pt\hbox{$>$}}}1.1 \times 10^{4}$ cm$^{-3}$. In the latter case, the nominal transit times of moving clouds indicate $r {\mathrel{\rlap{\lower4pt\hbox{\hskip1pt$\sim$}} \raise1pt\hbox{$<$}}} 0.9$ pc. It is not clear what properties of PG~0935+417 might produce this unusual outflow. The quasar is exceptionally luminous, with $L\sim6\times10^{47}$ ergs s$^{-1}$, but it has just a modest Eddington ratio, $L/L_{Edd}\sim 0.2$, and no apparent unusual properties compared to other quasars. In fact, PG~0935+417 has significantly less X-ray absorption than typical BAL quasars even though its outflow has a degree of ionization typical of BALs at speeds that are 2--3 times larger than most BALs.

Is PG0935+417 a rare or a typical case? With the goal of obtaining the frequency of every outflow class, including mini-BALs and outflows at extremely high speeds (up to $\sim$~0.2c), we created a systematic accounting of quasar outflows using the Sloan Digital Sky Survey (SDSS). We analyzed the $\sim$~2,200 highest signal-to-noise ratio spectra with emission redshifts between 1.6$<$z$<$3.5 to be able to observe high-velocity {{\rm C~}\kern 0.1em{\sc iv}~$\lambda\lambda 1548, 1550$} outflows given the SDSS spectral coverage (\citealt{RodriguezHidalgo09}; Rodr\'iguez Hidalgo et al. in prep). We found that mini-BALs are as frequent as BALs: they appear in $\sim$~11\% of SDSS quasars overall and in $\sim$~5\% of quasars without BALs.

To study dynamical and evolutionary effects, as well as to place constraints on the location of these outflows, we started a monitoring campaign of narrow, mini-BALs and borderline BALs, observed as {{\rm C~}\kern 0.1em{\sc iv}~$\lambda\lambda 1548, 1550$} absorption, selected from our SDSS sample. Here we present the preliminary results of the study of twenty six quasars observed during three observing runs: two with the GoldCam spectrometer at the 2.1 m telescope at Kitt Peak NOAO facilities and one with the Boller and Chivens CCD spectrograph at the MDM 2.4m telescope. When comparing this to the absorption features in the SDSS spectra we found that:

\begin{itemize}
\item 57\% of the quasar spectra included a {\hbox{{\rm C~}\kern 0.1em{\sc iv}}} absorption trough, of any class, that varied. 49\% out of the 39 studied absorption systems show variability. If we consider only our targeted mini-BALs and extremely high velocity outflows, 52\% of the absorption features varied. This agrees with the result found in a sample of BALQSOs where 65\% change between 2 observations (\citealt{Capellupo11}) 
\item Variability is complex and absorption profiles emerge, disappear, increase/decrease depths, but there are not clear signs of acceleration or deceleration
\item We observed that variability occurs over a range of $\Delta$t = 0.9--3.3 years in the quasar rest-frame 
\item Variability is found at all studied velocities: from $-5000$ to $-60000$ {\hbox{km~s$^{-1}$}}   . We did not find a significant trend between the velocity of the outflows and their variability
\item Narrow systems (NALs) seem to vary much less frequently. Most of these NALs might not form in quasar outflows and are likely to be unrelated to the quasar environment
\item Some systems change largely in depth and velocity range, and therefore cross our classification classes: BALs become mini-BALs and mini-BALs become BALs.
\end{itemize}

\section{The Transition from Mini-BAL to BAL in the quasar J115122+020426}
\label{sec3}

J115122+020426 (g=19.1) is a radio-quiet quasar observed for the first time as part of the SDSS first data release (\citealt{Schneider03}). In our survey of {\hbox{{\rm C~}\kern 0.1em{\sc iv}}} absorption in SDSS quasar spectra (Rodr\'iguez Hidalgo et al.~in prep, see \S\ref{sec2}) we found that the 2001 spectrum of J115122+020426 included, among others, an absorption feature at $\sim$~5108 {\AA} ($z_{abs}=$~2.296) that could be classified as a candidate {\hbox{{\rm C~}\kern 0.1em{\sc iv}}} mini-BAL with a FWHM of 1223 {\hbox{km~s$^{-1}$}}   . The width of the lines is broad enough that the two lines in the {\hbox{{\rm C~}\kern 0.1em{\sc iv}}} doublet are blended. We will refer to this absorption feature as system A.

\begin{figure}
\begin{center}
\includegraphics[width=10.5cm]{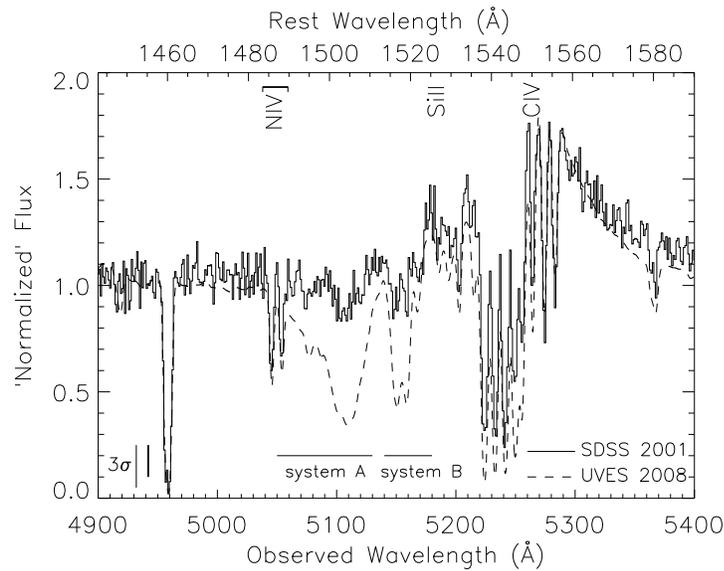}
\caption[Figure2]{SDSS (solid) and UVES (dashed) spectra of the quasar J115122+020426 taken in 2001 and 2008, respectively. The spectra are shown around the {\hbox{{\rm C~}\kern 0.1em{\sc iv}}} absorption troughs after dividing the spectra by a power law (in the case of the SDSS spectrum) and a second order polynomial (in the UVES spectrum) to show only variations in absorption. Emission lines are labeled at the top of the spectra. The UVES spectrum has been smoothed to match the SDSS resolution. Horizontal lines indicate the {\hbox{{\rm C~}\kern 0.1em{\sc iv}}} absorption features of interest: systems A ($\sim$~5100 {\AA}) and B ($\sim$~5150 {\AA}). Typical error sizes are represented on the bottom left corner. Both absorption features appear to be much weaker in the SDSS spectrum than in the UVES one. System A transitions from a mini-BAL to a BAL.}
\label{fig2}
\end{center}
\end{figure}

After comparison of the SDSS spectrum to archival spectra taken at the Very Large Telescope with the Ultraviolet and Visual Echelle Spectrograph (VLT/UVES), we notice that this absorption trough transitions from the {\hbox{{\rm C~}\kern 0.1em{\sc iv}}} mini-BAL into a BAL. Figure \ref{fig2} shows both spectra: the solid line represents the SDSS 2001 spectrum and the dashed line the VLT/UVES 2008 spectrum. For easy visualization we plot both spectra after a basic ``normalization'': a power law (in the case of SDSS) and a second order polynomial (in the case of UVES), using regions of the wavelength coverage devoid of both absorption and emission. Also, for this figure, we smooth the UVES spectrum to the SDSS resolution. The rest-frame wavelengths are defined relative to the redshift $z_{em} =$~2.396, measured from the uncontaminated centroid of the {\hbox{{\rm C~}\kern 0.1em{\sc iii}}}$\lambda$1909 emission line.
Together with system A, another absorption system at $\sim$~5150 {\AA} ($z_{abs}=$~2.329; hereafter, system B) also appears to have significantly increased in strength between the two epochs. This system is outflowing at a speed of $v\approx -6000$ {\hbox{km~s$^{-1}$}}    relative to the quasar. The vertical lines in the bottom left corner represent the 3{$\sigma$} errors in the SDSS and VLT/UVES spectra. Some residuals of the {\hbox{{\rm C~}\kern 0.1em{\sc iv}}} emission line are likely to be present (see the region around $\sim$~5270 {\AA}).

From Fig. \ref{fig2} we can see that the variability is substantial: the rest-frame equivalent width has increased in $\sim$~8 times in system A and $\sim$~2.5 times in system B. The deepest parts of the trough in the SDSS spectrum remain the strongest parts of the absorption in the UVES spectrum, and regions that were unabsorbed in the SDSS spectrum appear absorbed in the UVES one. This is typical of the other variations we have observed in our monitoring program and a characteristic that needs to be included in theoretical models to understand these absorbers and what is producing their variability. We also note that the narrow absorption lines at lower velocities seem to increase in strength like the broader features, indicating that these lines are also intrinsic to the quasar and might be part of the same outflow as the broad lines.

\acknowledgements This work was supported by the Alumni Fellowship at UF, the Chandra award GO1-12146C and the XMM-Newton program number 50462.

\bibliographystyle{asp2010}
\bibliography{bibliography_proc.bib}

\end{document}